# Strictly One-Dimensional Electron System in Au Chains on Ge(001) Revealed By Photoelectron K-Space Mapping

S. Meyer[1], J. Schäfer[1], C. Blumenstein[1], P. Höpfner[1], A. Bostwick[2], J.L. McChesney[2], E. Rotenberg[2] and R. Claessen[1]

[1]*Physikalisches Institut, Universität Würzburg, 97074 Würzburg, Germany*
[2]*Advanced Light Source, Lawrence Berkeley National Laboratory, Berkeley 94720, California, USA*

Atomic nanowires formed by Au on Ge(001) are scrutinized for the band topology of the conduction electron system by k-resolved photoemission. Two metallic electron pockets are observed. Their Fermi surface sheets form straight lines without undulations perpendicular to the chains within experimental uncertainty. The electrons hence emerge as strictly confined to one dimension. Moreover, the system is stable against a Peierls distortion down to 10 K, lending itself for studies of the spectral function. Indications for unusually low spectral weight at the chemical potential are discussed.

Self-organized atomic chains on semiconductors offer a variety of architectures, and provide model systems to study exotic physics in a nearly one-dimensional (1D) situation. Representatives include In chains on Si(111) [1], Au on Si(557) and (553) [2,3], and the recent addition of Au chains on Ge(001) [4]. Such chains with tunable properties are in focus as critical test of predictions from solid state theory. On one hand, this pertains to the Peierls instability, where a charge density wave (CDW) leads to a metal-insulator transition [5]. Secondly, for strictly 1D systems, a novel correlated electron state referred to as Luttinger liquid (LL) [6] is proposed, with collective excitations of spin and charge.

The key technique to address these issues is angle-resolved photoemission (ARPES). As paramount criterion for a CDW, a nesting condition must exist in a Fermi surface of quasi-1D character, accompanied by energy gaps. In contrast, in a LL picture the spectral weight at low temperature vanishes smoothly towards the chemical potential μ without a gap. CDW formation was reported, e.g., for In/Si(111) [1] already at moderate cooling to ~200 K [1,7]. Since fluctuations hinder an idealized 1D system from ordering [5], significant inter-chain coupling is required to stabilize the Peierls phase. In ARPES, a half-filled band with quasi-1D character is identified [1]. The necessary higher-dimensional coupling is enhanced by two additional metallic bands with noticeable lateral dispersion, which are also affected by the gap opening [8]. Hence, a realistic situation with multiple bands drastically modifies a naive Peierls picture, and seemingly inhibits a Luttinger regime.

In the search for Luttinger physics, vanishing spectral weight was seen, e.g., in seminal ARPES studies on $Li_{0.9}Mo_6O_{17}$ [9]. A LL claim was also made for Au/Si(557) nanowires [2], based on an alleged splitting of the dispersion into spin and charge branches. Careful ARPES work subsequently excluded this by revealing two conventional electron bands [10]. Moreover, small energy gaps are observed [11], pointing at a Peierls-type instability while ruling out a Luttinger state [6]. Thus, for atomic chains this thrust is still ongoing. In this respect, the noble metal chains on Ge(001) attract attention [4,12,13]. The recently discovered Au chains [4] have a uniquely elevated architecture and a pronounced spatial separation, suggestive of low inter-chain coupling. This spurs heightened interest in the resulting degree of electronic 1D character. STM investigations can provide at best a qualitative indication [14]. Yet, except for ambiguous low-statistics data [15], high-resolution ARPES Fermi surface mapping to address the dimensionality, manifold bands, nesting conditions, and potential energy gaps versus unusual line shapes is missing to date.

In this Letter, we report on ARPES k-space mapping of these chains at 15 K, which reveals a novel band topology that significantly differs from all previous nanowire systems. It represents the first *single band* system, with two electron pockets on either side of the Brillouin zone. The Fermi surface consists of *parallel lines*, with possible variations from 1D behavior below detection limit. The incommensurate band filling does not support a simple Peierls scenario, and a CDW is absent in electron diffraction down to 10 K. Instead, appreciable deviations from Fermi liquid behavior are observed in the 1D band near the chemical potential.

Experimentally, Au was deposited onto n-doped Ge(001) at a substrate temperature of ~500 °C, which induces self-organized nanowire growth [4]. ARPES at 15 K was performed at the Advanced Light Source at beamline 7.0.1. The total energy resolution with a Scienta R4000 analyzer was set to ~30 meV at a photon energy of hν = 100 eV.

A salient feature of the nanowires, seen in STM Fig. 1(a), is that they are raised above the substrate [16]. The lateral spacing amounts to 1.6 nm. Two orientation domains (rotated by 90°) result from terrace steps of the substrate. The Au/Ge(001) chains lend themselves to ARPES studies because they exhibit excellent long-

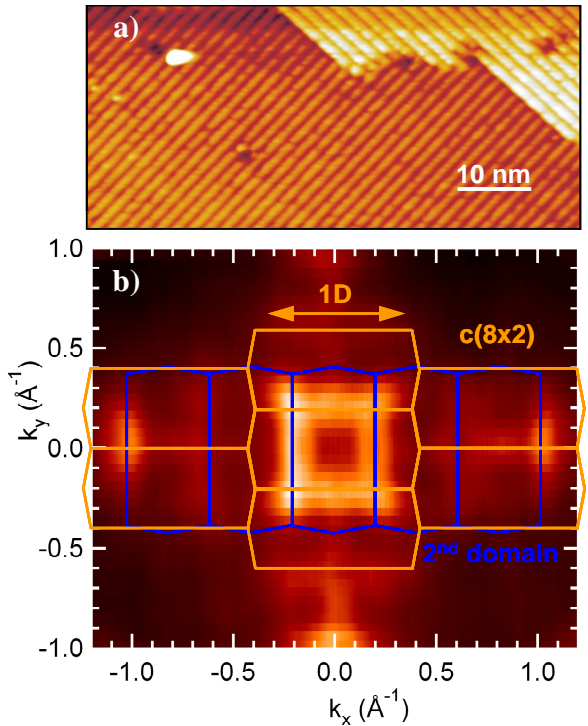

FIG. 1 (color online). a) STM image of the Au-induced nanowires (empty states, bias +1.0 V, 1.0 nA), showing two domains from terrace stacking. b) ARPES Fermi surface overview at hν = 100 eV, T = 15 K from dual-domain sample. Two sheets within the SBZ are rather straight perpendicular to the corresponding 1D direction. Higher SBZs appear suppressed, ascribed to transition matrix element effects.

range order, reflected in their c(8×2) low-energy electron diffraction (LEED) pattern [4].

An overview of the constant energy surface at $E_F$, commonly referred to as "Fermi surface", is presented in Fig. 1(b). Note that the square structure solely results from the dual domains; nonetheless the respective sheets are well separated from their rotated counterparts. The c(8×2) surface Brillouin zone (SBZ) in Fig. 1(b) is a stretched hexagon of ± 0.42 Å$^{-1}$ extent on the 1D axis to the zone boundary (ZB). Photoemission intensities are remarkably suppressed in higher SBZs, which we ascribe to optical transition matrix element effects which can cause significant modulations [17]. Already in this coarse-grid overview one observes a strikingly linear shape of the Fermi surface sheets, to be studied in detail below. This clearly disproves an earlier interpretation as a two-dimensional metallic state [15].

The central cut through $\bar{\Gamma}$ in Fig. 2(a) shows the band situation along the chains. By variation of photon energy the surface character of the bands is identified, and at hν = 100 eV the Ge bulk bands are tuned away. Two shallow electron pockets are located on either side of $\bar{\Gamma}$. Faint intensity is observed below them at higher binding energies, relating to deeper lying bands. The band situation near the Fermi surface of a single nanowire domain is sketched in Fig. 2(b), consisting of two troughs formed by the electron pockets.

These bands are of roughly symmetric shape with a band minimum at ~0.20 Å$^{-1}$, located approximately at half the distance to the ZB. No such surface states are known from bare Ge(001) [18], so they must originate from the chain reconstruction. The Fermi level crossings on either side of $\bar{\Gamma}$ are spaced by ~ 0.12 Å$^{-1}$. The occupied band width is ~100 meV, which is small compared to the ~500 meV range for In/Si(111) [8] and the eV-range for Au on stepped silicon [19].

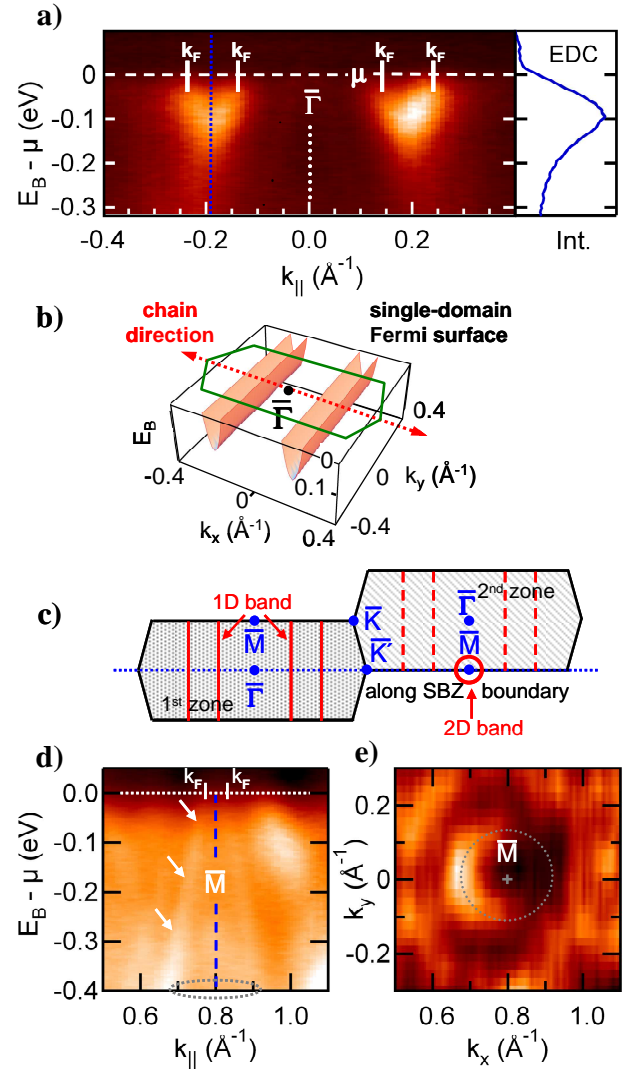

FIG. 2 (color online). a) ARPES band map at 100 eV and T = 15 K along chain direction through $\bar{\Gamma}$. Two shallow electron pockets are seen (energy distribution curve (EDC) taken at blue dotted line). b) Schematic of band situation for single-domain Au chains on Ge(001), consisting of two troughs. c) Schematic of SBZ alignment. d) Band map along $\bar{K}'-\bar{M}$ of 2nd SBZ, showing a metallic hole band at $\bar{M}$ used as intrinsic reference for μ. e) Constant energy surface at 0.4 eV binding energy, revealing the 2D character of the band at $\bar{M}$.

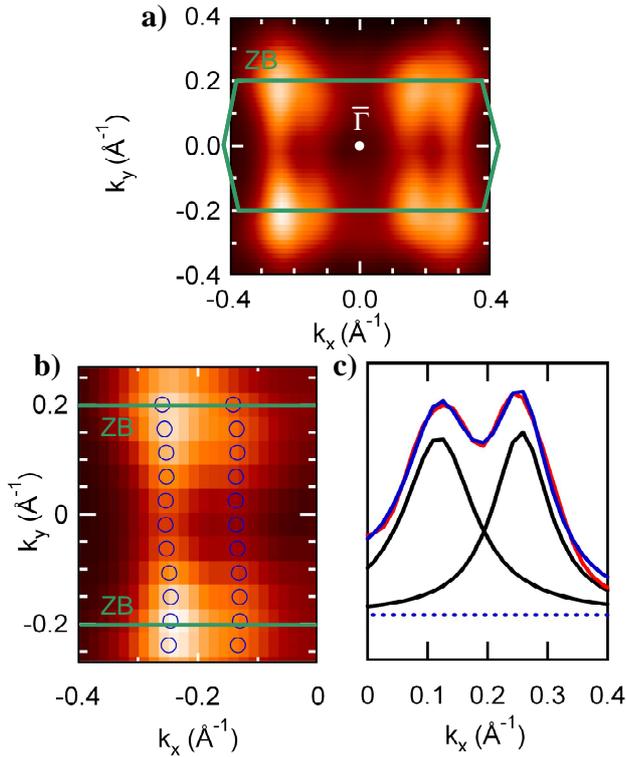

FIG. 3 (color online). a) Fermi surface (energy window $\mu$ to $\mu - 30$ meV, to compensate for low spectral weight at $\mu$) of sample with domain imbalance. b) Close-up Fermi surface data (window $\mu$ to $\mu - 30$ meV) with $k_F$-positions from MDC-fits (blue circles). Variations in the perpendicular $k_F$-lines are absent except for statistical uncertainty. c) MDC-fit for line through $\overline{\Gamma}$, showing two clearly separated $k_F$'s.

In passing on the 1D axis into the 2nd SBZ, the ARPES scan coincides with the SBZ boundary line through $\overline{M}$, as illustrated in Fig. 2(c). The ARPES data in Fig. 2(d) reveal a parabolic hole band at $\overline{M}$ with weak intensity. A two-dimensional (2D) character is determined from a constant energy surface, Fig. 2(e), hence it must obey conventional Fermi liquid statistics. This 2D band forms a small hole pocket and exhibits a metallic Fermi edge (see spectra in Fig. 4(d) below). This *intrinsic* reference is used to precisely determine the *chemical potential* $\mu$ (usually termed Fermi level $E_F$). This is superior to a metal foil reference due to possible surface photo voltage of the sample.

It is important to note that the 2D hole band is *not* located directly at the surface. This follows from the tunneling spectroscopy data in [4], where a 2D DOS (which is constant upon energy) would lead to a step in the spectrum just above the Fermi level, yet which is not observed. Thus, the band must be localized in subsurface layers. Such substrate-derived states with hole-like dispersion are well known to occur as interface states in metal-adsorbed Ge (in addition to the adatom surface states), and are located within ~ 10–20 layers below the topmost layer [20,21].

Precise evaluation of the dimensionality is based on k-space mapping of the constant energy contours near $\mu$, which host the 1D electron liquid. High-statistics Fermi surface data for samples with a domain imbalance (relating to a small incidental crystal miscut), as in Fig. 3(a), can be used to trace the dispersion. On either side of $\overline{\Gamma}$ two ridges corresponding to the Fermi vectors of the electron pocket are identified. The resulting Fermi surface lines appear very straight throughout, notably extending up to the zone boundary. The band filling $f$ is obtained from its relation to the SBZ area as fraction of occupied $k_x$-range, $f \sim 0.12$ Å$^{-1}$/ 0.80 Å$^{-1}$ ~0.15 for each pocket, yielding a total band filling of ~0.30.

Quantitative dispersion evaluation must rely on the momentum distribution curves (MDCs). Although the signal at $\mu$ is largely suppressed in a LL scenario discussed below, it contains sufficient intensity from energy resolution broadening. The MDC peaks in the close-up of the Fermi surface, Fig. 3(b), are fitted with Lorentzians to determine the $k_F$'s. The MDC maxima are well separated, as in the example in Fig. 3(c). This ana-

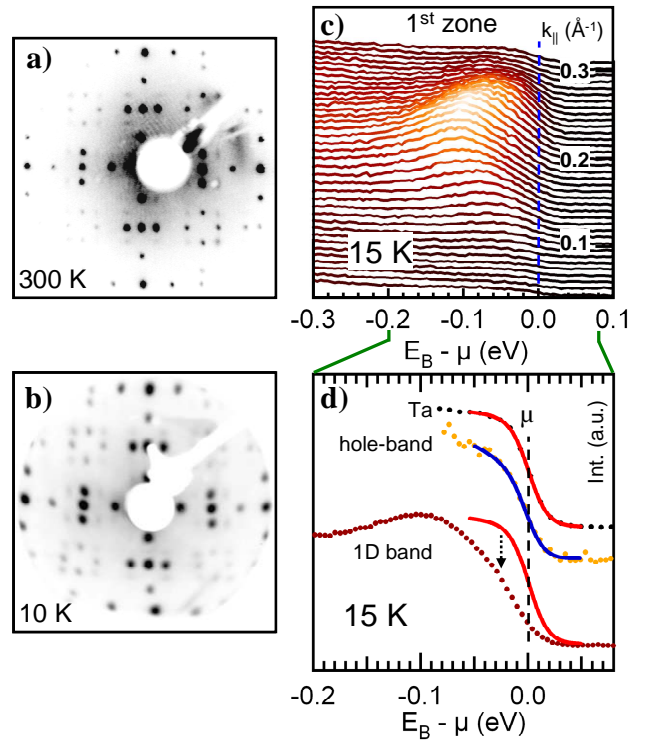

FIG. 4 (color online). a) LEED pattern (21 eV) of c(8×2) reconstruction (dual domain) at 300 K. b) LEED pattern (24 eV) at 10 K. It shows no additional spots so that a CDW is excluded. c) EDCs of the 1D electron pocket at 15 K (1st SBZ). d) EDCs at 15 K of Ta-foil (aligned with $\mu$) and metallic 2D band used for $\mu$-determination, fitted with resolution-broadened Fermi function including a linearly sloping density of states (red and blue). The angle-integrated EDC of the 1D band (0.10-0.30 Å$^{-1}$) deviates from the metallic Fermi edges (red: Ta Fermi fit overlay), exhibiting suppressed spectral weight within ~50 meV below $\mu$.

lysis of the dispersion perpendicular to the chains confirms that it is virtually perfectly straight. Remaining small scatter in the $k_F$-positions reflects the statistical uncertainty of the MDC analysis of the two closely adjacent peaks. For each $k_F$ contour line the accuracy of the analysis amounts to $\delta k = 0.004$ Å$^{-1}$, determined as standard deviation of 110 independent MDC fits of several data sets. This value provides a tight upper boundary for possible undulations within the Fermi surface.

In contrast, significantly larger Fermi surface curvatures are reported for, e.g., Au on Si(553) (undulation $\delta k \sim 0.03$ Å$^{-1}$) which indicates a substantial coupling between neighboring wires [3,19]. Notably, both Au-Si(553) and Au-Si(557) exhibit a complex multiband situation, and all these stepped systems exhibit lattice instabilities upon cooling. The In chains on Si(111), likewise with a phase transition, also show multiple bands at the Fermi surface and a noticeable dispersion for the most 1D-like band, which is attributed to interband interaction [8].

The shape of the Fermi contour is suggestive of various *nesting conditions* for a CDW, with distortion vectors $q_1 = 0.12$ Å$^{-1}$ ~0.15 G, $q_2 = 0.28$ Å$^{-1}$ ~0.35 G, and $q_3 = 0.52$ Å$^{-1}$ ~0.65 G (where G = 0.80 Å$^{-1}$ is the reciprocal translation vector in 1D direction). None of them appears to be commensurate with the lattice.

The occurrence of a CDW superstructure can best be probed with temperature-dependent LEED. The LEED pattern in Fig. 4(a) at 300 K reflects the c(8×2) long-range order. However, when going to much lower temperatures, no additional superstructure is observed, as evident from the LEED data at 10 K in Fig. 4(b). Hence formation of a CDW that would imply energy gap formation can be excluded. This opens the pathway to study the non-perturbed low-energy spectral function of the 1D electron band at low temperature with respect to non-Fermi liquid physics.

In the energy distribution curves (EDCs) of the 1D band at 15 K in Fig. 4(c) it becomes apparent that the spectral intensity is rather low near µ. This can be compared to the Fermi edge of the Ta clip in Fig. 4(d). It reflects the experimental resolution (~30 meV) and the small thermal broadening (4kT ~5 meV). The metallic 2D band at $\overline{M}$ (used as intrinsic reference for µ) closely replicates the Ta Fermi edge, albeit with slightly sloping background. In addition, Fig. 4(d) shows the angle-integrated spectrum covering the whole range of the 1D electron pocket. In contrast to the Fermi edges of Ta and the 2D band, the data at 15 K indicate that the spectral weight is experiencing a suppression in a range of ~ 50 meV below µ.

Since our structural investigations exclude a static Peierls distortion (which would imply an energy gap), one needs to consider *non-Fermi liquid* behavior. For 1D electron systems described by LL theory, the spectral weight of the angle-integrated density of states is predicted to vanish at µ with a power-law behavior upon energy (where the exponent reflects the interaction strength) [6]. Experimentally this has been found, e.g., in the lithium purple bronze [9] and in carbon nanotubes [22], while no such report exists for atomic nanowires. The present data now show that the intensity is substantially suppressed at µ, with remaining intensity ascribed to the temperature (15 K) and to resolution broadening. Here we refrain from further analysis since the resolution is comparable to the energy scales for LL physics. Quantitative line shape modeling will require dedicated ARPES studies at very high resolution, which must also probe the characteristic temperature behavior [23].

In conclusion, the electron system of the Au/Ge(001) chains mapped by ARPES exhibits straight Fermi surface lines, whose 1D character is exceptionally high. At the same time, a CDW can be excluded from LEED even at 10 K. As the only nanowire system, its 1D electron topology thus satisfies the requirements for LL physics, and first indications for non-Fermi liquid behavior are seen. The two electron pockets, each with ~1/8 filling, call for corresponding theoretical modeling of such 1D chains.

The authors are grateful to Y. S. Kim, L. Patthey and T. Umbach for technical support, and funding by the DFG (Scha 1510/2-1 and FOR 1162) and DOE (DE-AC03-76SF00098).